%% file: qpace4.tex
\documentclass[10pt, conference, compsocconf]{IEEEtran}

\usepackage{graphicx}
\usepackage{amsmath}
\usepackage{url}
\usepackage{units}
\usepackage{listings}

\DeclareMathOperator{\re}{Re}
\DeclareMathOperator{\im}{Im}
\newcommand{\I}{i}

\begin{document}

\title{SVE-enabling Lattice QCD Codes\\ {\large Workshop paper: REV-A 2018}}
\author{\IEEEauthorblockN{
  Nils Meyer\IEEEauthorrefmark{1},
  Peter Georg\IEEEauthorrefmark{1},
  Dirk Pleiter\IEEEauthorrefmark{1}\IEEEauthorrefmark{2},
  Stefan Solbrig\IEEEauthorrefmark{1}, and
  Tilo Wettig\IEEEauthorrefmark{1}
}
\IEEEauthorblockA{\IEEEauthorrefmark{1}%
  University of Regensburg, 93040 Regensburg Germany;
  \url{{nils.meyer,peter.georg,stefan.solbrig,tilo.wettig}@ur.de}
}
\IEEEauthorblockA{\IEEEauthorrefmark{2}%
  Forschungszentrum J\"{u}lich, 52425 J\"{u}lich, Germany;
  \url{d.pleiter@fz-juelich.de}
}
}

\IEEEoverridecommandlockouts%
\IEEEpubid{\makebox[\columnwidth]{%
\parbox{\columnwidth}{IEEE International Conference on Cluster Computing,%
\newline September 10-13, 2018, Belfast, UK%
\newline \copyright 2018 IEEE, DOI: 10.1109/CLUSTER.2018.00079 \hfill}}
\hspace{\columnsep}\makebox[\columnwidth]{ }}

\maketitle

\IEEEpubidadjcol

\begin{abstract}
Optimization of applications for supercomputers of
the highest performance class requires parallelization at multiple
levels using different techniques. In this contribution we focus on
parallelization of particle physics simulations through vector
instructions. With the advent of the Scalable Vector Extension (SVE)
ISA, future ARM-based processors are expected to provide a
significant level of parallelism at this level.
\end{abstract}


\input{section_introduction}
\input{section_lqcd_and_grid}
\input{section_arm_sve}
\input{section_code_examples}
\input{section_sve_enabling_grid}
\input{section_related_work}
\input{section_conclusion}

% ==================================================================================================
\section*{Acknowledgment}
% ==================================================================================================

We acknowledge the funding of the QPACE4 project provided by the Deutsche
Forschungsgemeinschaft (DFG) in the framework of SFB/TRR-55.
Furthermore, we acknowledge the support from the HPC tools team at ARM, in particular
Ashok Bhat, Juan Gao, Assad Hashmi, and Will Lovett.

% ==================================================================================================
\bibliographystyle{IEEEtran}
\bibliography{qpace4}
% ==================================================================================================

\end{document}

%% file: section_introduction.tex
% ==================================================================================================
\section{Introduction}
% ==================================================================================================

% LQCD introduction
Understanding the nature of the strong interactions, one of the four fundamental interactions in physics,
is still an important challenge.
Quantum Chromodynamics (QCD), the theory believed to describe these interactions,
was already established in the 1970s.
For many cases large-scale numerical simulations are needed to study QCD.
To facilitate such simulations the theory is formulated in a discretized and
computer-friendly version called Lattice QCD (LQCD) \cite{PhysRevD.10.2445}.
For state-of-the-art research in LQCD supercomputers with a throughput of \unit[O(10)]{PFlop/s}
are used.

% Community software
With the architectures for such high-end supercomputers becoming more parallel and complex,
the design of software exploiting these machines has become a challenge for researchers in
the field of LQCD.
To share the burden of designing highly scalable applications, several efforts on creating
community codes have been started to provide libraries that can be used for applications
of different research groups.
Examples are the Chroma software system \cite{Edwards:2004sx} or
the QUDA library \cite{Clark:2009wm}.
In this contribution, we focus on a recent effort called Grid \cite{2015arXiv151203487B}.

% SVE
Grid is a framework for LQCD simulations that was designed for processor architectures 
featuring very wide SIMD instructions, such as AVX-512, a 512-bit wide ISA for x86 architectures.
Future ARM-based processor architectures will support a vector ISA called
Scalable Vector Extension (SVE) \cite{7924233}, which would allow for vectors of length up
to \unit[2048]{bits}.

% This contribution
In this paper we make the following contributions:
\begin{enumerate}
\item We propose a strategy for porting the LQCD framework Grid to the
      SVE ISA and report on first experiences using the current development toolchain
      and available emulators.
\item We discuss and analyze different ways of
      implementing complex arithmetics exploiting the SVE ISA.
\item Based on different porting strategies we analyze and demonstrate that the SVE ISA allows
      for an efficient implementation of key computational patterns used in LQCD applications.
\end{enumerate}

% Paper structure
This paper is organized as follows:
In Section~\ref{sec:lqcd} we provide a brief introduction to LQCD, its main numerical kernel
and the domain-specific software framework considered in this paper, namely Grid.
In Section~\ref{sec:sve} we highlight the most important features of SVE, and in Section~\ref{sec:sve_example} we provide various SVE code examples.
In Section~\ref{sec:sve-enabling-grid} we document our strategy for porting Grid and
report in Section~\ref{sec:verification} on how we verified our SVE-enabled version thereof.
In Section~\ref{sec:related-work} we provide a brief overview of
related work before concluding in Section~\ref{sec:conclusion}.

%%% Local Variables:
%%% mode: latex
%%% TeX-master: "qpace4"
%%% End:

%% file: section_lqcd_and_grid.tex
% ==================================================================================================
\section{LQCD Simulations and Grid}\label{sec:lqcd}
% ==================================================================================================

% --------------------------------------------------------------------------------------------------
\subsection{Overview}
% --------------------------------------------------------------------------------------------------

% LQCD solver, matrix-vector product
A significant fraction of time-to-solution of LQCD applications is spent in solving a
linear set of equations, for which iterative solvers like Conjugate Gradient are used.
The most compute-intensive task typically is the product of the lattice Dirac
operator and a quark field $\psi$.\footnote{In the following we focus on a particular
formulation of LQCD using so-called Wilson fermions.}
A quark field $\psi^{ia}_x$ is defined
at lattice site $x = 0, \ldots, V-1$ and carries so-called color indices $a = 1, \dots,
3$ and spinor indices $i = 1, \dots, 4$. Thus, $\psi$ is a vector with $12\,V$ complex entries.
$V$ denotes the size of the 4-dimensional lattice.
Today's state-of-the-art simulations use lattices with a minimal size of
$V = L^3\times T=32^3\times 64$.
The so-called hopping term $D_h$ of the Dirac operator acts on $\psi$ as follows:
\begin{align}
  \psi^\prime_x &= D_h \psi   \label{eq:Dirac} \\ \notag
                &= \sum_{\mu=1}^4\left\{
    U_{x,\mu}(1+\gamma_\mu)\psi_{x+\hat\mu} +
    U^\dagger_{x-\hat\mu,\mu}(1-\gamma_\mu)\psi_{x-\hat\mu}\right\}.
\end{align}
Here, $\mu$ labels the four space-time directions, and $U_{x,\mu}$ are
the SU(3) gauge matrices associated with the links between nearest-neighbor lattice sites.
The gauge matrices carry color indices and are represented by $3 \times 3$ matrices with 
complex entries.
The $\gamma_\mu$ are the (constant) Dirac matrices, carrying spinor indices.

% Parallelization
Parallelization of the matrix-vector product is achieved by a domain decomposition in 1 to 4
dimensions.
Therefore, the larger the lattice size $V$ the higher the intrinsic level of parallelism.
In particular, due to the regular structure of the problem, parallelization can be performed
at multiple levels.
For the coarsest level a set of sub-lattices is distributed over (a very large number of)
different processes, e.g., different MPI ranks.
Further parallelization within a process is achieved through thread-level parallelization,
e.g., using OpenMP, as well as through vectorization at the instruction level.

% --------------------------------------------------------------------------------------------------
\subsection{Data Layout}\label{sec:dlayout}
% --------------------------------------------------------------------------------------------------

\begin{figure}[bht]
\begin{center}
\includegraphics[width=.95\columnwidth]{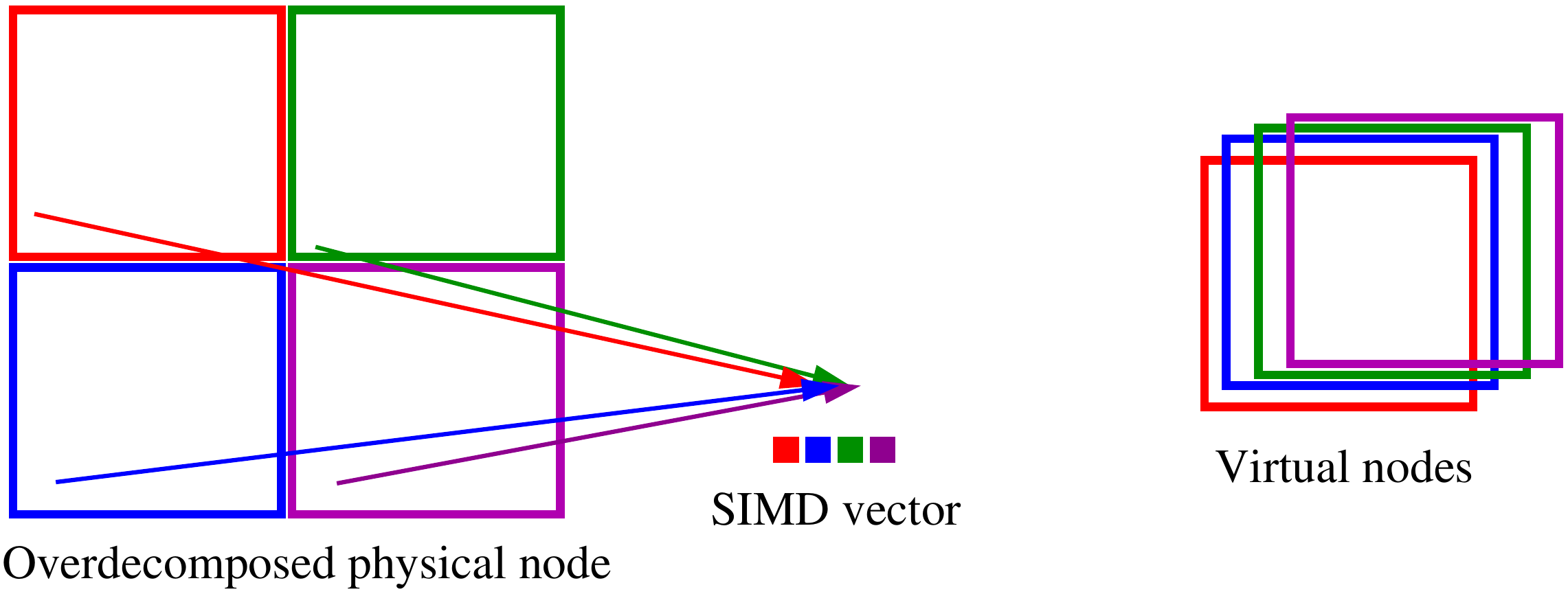}
\end{center}
\caption{\label{fig:overdecompose}Decomposing a sub-lattice over multiple virtual nodes
\cite{2015arXiv151203487B}.}
\end{figure}

% Data layout
For implementing parallelization the choice of a suitable data layout is crucial.
For instance, in case of vectorization a distribution of data such that data for
neighboring lattice sites are distributed over a single vector results in the need for
combining different elements of the same vector for a single application of the hopping
term as defined in Eq.~\eqref{eq:Dirac}.
To address this problem, Grid implements the concept of ``virtual nodes.''
Within a single thread the sub-lattice is distributed over a set of such virtual nodes
as shown in Fig.~\ref{fig:overdecompose},
where the number of virtual nodes per thread is typically equal to the number of vector elements.
By keeping the size of the sub-lattice processed by a single virtual node sufficiently large,
neighboring lattice sites will be assigned to different vectors.

% --------------------------------------------------------------------------------------------------
\subsection{Architecture-Specific Implementations}\label{sec:grid}
% --------------------------------------------------------------------------------------------------

% Hiding data layout in Grid
Grid is designed to maximize the flexibility in choosing the data layout
optimal for parallelizing on a given architecture without compromising on
portability. By implementing a suitable abstraction layer based on C++
template expressions, the complexity is hidden from the user.

% ISA-specific implementation
The software is organized such that the machine-specific aspects are
confined to a small number of lines of code in a small number of files.
Inline assembly is used for the implementation of the
Dirac operator in Eq.~\eqref{eq:Dirac} for some instruction-set architectures, e.g., AVX-512.
The assembly is optimized for register and cache reuse. All the other
machine-specific code is implemented using intrinsics, including:
\begin{itemize}
\item arithmetics of real and complex numbers,
\item permutations of vector elements,
\item load, store, memory prefetch, streaming memory access,
\item conversion of floating-point precision.
\end{itemize}
Machine-specific implementations exist for a variety of Intel architectures,
ARM NEONv8 and also IBM BlueGene/Q. Table~\ref{tab:gridsupportedarchitectures}
shows the architectures supported by Grid at the time of writing
this contribution.

\begin{table}[bht]
\caption{Architectures supported by Grid.}
\begin{center}
\begin{tabular}{l|l}
    SIMD family        & Vector length \\\hline\hline
    Intel SSE4         & 128 bit \\
    Intel AVX/AVX2     & 256 bit \\
    Intel ICMI, AVX-512 & 512 bit \\\hline
    IBM QPX            & 256 bit \\\hline
    ARM NEONv8         & 128 bit \\\hline
    generic C/C++      & architecture independent,\\
                       & user-defined array size
\end{tabular}
\end{center}
\label{tab:gridsupportedarchitectures}
\end{table}

%%% Local Variables:
%%% mode: latex
%%% TeX-master: "qpace4"
%%% End:

%% file: section_arm_sve.tex
% ==================================================================================================
\section{ARM Scalable Vector Extension}\label{sec:sve}
% ==================================================================================================

The ARM Scalable Vector Extension (SVE) is a novel vector extension for ARMv8
architectures. The SVE ISA facilitates a significantly higher single-core performance
and thus targets applications with high demand for computational power,
such as high-performance computing but also machine learning~\cite{7924233}.

% -------------------------------------------------------------------------------------------------
\subsection{Features of SVE}
% -------------------------------------------------------------------------------------------------

% Features of the SVE ISA
For a comprehensive list of features of SVE we refer
to \cite{7924233}. Here we list the features of the SVE ISA
we believe to be beneficial for LQCD applications:
\begin{itemize}
\item wide vector units,
\item structure load/store instructions supporting load/store of an array of $n$-element
structures into $n$ vectors, with one vector per structure element,
\item vectorized 16-, 32-, 64-bit floating-point operations,
  including arithmetic operations and conversion of precision,
\item vectorized arithmetic of complex numbers.
\end{itemize}

% SVE ACLE
Convenient access to features of SIMD extensions is typically provided by
intrinsics, i.e., built-in functions handled specially by the compiler.
The ARM C Language Extensions (ACLE) for SVE intrinsics provide access
to features of the SVE hardware in C/C++~\cite{acle}.

% --------------------------------------------------------------------------------------------------
\subsection{Vector-Length Agnosticism}
% --------------------------------------------------------------------------------------------------

% SVE and VLA
SVE does not define the size of the vector registers, but
constrains it to a range of possible values, from a minimum of 128 bits up
to a maximum of 2048 in multiples of 128. The silicon provider chooses
the vector-register length and defines the performance characteristics of the
hardware.
SVE pursues a so-called Vector-Length Agnostic (VLA) programming model that
allows code execution to dynamically adapt to the available vector length
at runtime. To achieve this SVE implements predication registers for masking
vector lanes for operations on partial vectors.

% -------------------------------------------------------------------------------------------------
\subsection{Limitations on Usage of SVE ACLE Data Types}
\label{sec:svelimitations}
% -------------------------------------------------------------------------------------------------

The SVE vector length is unknown at compile time.\footnote{
This statement is valid for the ARM clang SVE compiler. Other (commercial
or future) SVE compilers might be aware of the SVE vector length.}
Therefore SVE ACLE data types do not have a defined size.
A comprehensive list on the usage of these data types, also referred to
as ''sizeless structs,'' is provided in \cite{acle}.
In the following we restrict ourselves to limitations on the usage of
ACLE relevant for enabling SVE in Grid. SVE ACLE data types may not
be used:
\begin{itemize}
\item as data members of unions, structures, and classes,
\item to declare or define a static or thread-local storage variable,
\item as the argument to \texttt{sizeof}.
\end{itemize}
Implications of and solutions for these limitations are discussed in
Section~\ref{sec:proposals}.

% -------------------------------------------------------------------------------------------------
\subsection{Complex Arithmetic}
\label{sec:complexarith}
% -------------------------------------------------------------------------------------------------

The SVE ISA supports vectorized arithmetic of complex numbers
for 16-bit, 32-bit, and 64-bit floating-point data types.
This feature is of interest since complex multiplications, depending on
the data layout chosen, may require combining different elements of the
same vector. Without specific support for complex arithmetics additional
instructions may be required to re-order the vector elements.

Let $x, y, z$ denote vectors of complex numbers and $\I$ the
imaginary unit. SVE complex arithmetic includes:
\begin{itemize}
\item vectorized add/sub of complex numbers,
\begin{equation*}
  x_i \pm \I \, y_i  \,,
\end{equation*}
\item vectorized fused multiply-add/sub of complex numbers,
\begin{equation*}
  z_i \pm ( \re \, x_i ) \times y_i \,,
    \quad z_i \pm ( \I \im \, x_i ) \times y_i \,.
\end{equation*}
\end{itemize}
In this contribution we focus on complex multiply-add/sub.
The \texttt{FCMLA} instruction takes three vector registers as parameters,
with the real components in even elements and the imaginary
components in odd elements.
A fourth (immediate) parameter specifies discrete rotation
of the second input vector in the complex plane.
We refer to the SVE ACLE
specification for details on the usage
of the \texttt{FCMLA} instruction~\cite{acle}. For example,
the following complex calculations are enabled by
concatenating two \texttt{FCMLA} instructions each
(the asterisk denotes complex conjugation):
\begin{align}\label{eq:fcmla}
    z_i &\pm x_i \times y_i\,, \\
    z_i &\pm (x_i^*) \times y_i \,. \notag
\end{align}
Complex multiplication is achieved by setting $z_i=0$ in
Eq.~\eqref{eq:fcmla}.

%%% Local Variables:
%%% mode: latex
%%% TeX-master: "qpace4"
%%% End:

%% file: section_code_examples.tex
% ==================================================================================================
\section{SVE Code Examples}\label{sec:sve_example}
% ==================================================================================================

In this section we present simple C++ code examples and illustrate
the binary code generated by the ARM clang SVE compiler. We also
show how to exploit the SVE ISA for complex arithmetics using
ACLE. The examples are relevant for Grid.

We used the ARM armclang 18.3 SVE compiler to generate the binaries.
The compiler is based on clang/LLVM 5.0.1.
For the compilation process we used optimization level \emph{Ofast}
and defined the target architecture as \emph{arch=armv8-a+sve}. The
settings enable auto-vectorization for SVE. The armclang 18 compiler
is not aware of the hardware implementation of the vector length
and optimizes following the VLA paradigm.

For verification of the SVE binary code we used the ARM instruction
emulator (ArmIE) 18.1. The emulator allows for functional code verification
by emulating SVE instructions on AArch64 platforms. The SVE vector length
is supplied to ArmIE as a command-line parameter. We tested our examples
emulating multiple vector lengths.

% -------------------------------------------------------------------------------------------------
\subsection{Real Arithmetics}
\label{sec:cppcoderealarith}
% -------------------------------------------------------------------------------------------------

As the first example of SVE we consider pairwise multiplication of
array elements, with the result being stored in a third array.
The following code shows the C++ implementation of the
operation $z_i \leftarrow x_i \times y_i$ for arrays of the data type
\texttt{double} with $n$ elements each:

\lstinputlisting[
        language=C++,
        numbers=left,
        basicstyle=\tiny,
        numberstyle=\tiny,
        linewidth=1.0\textwidth,
        xleftmargin=0.03\textwidth
]{
code/vector_mult_real.cc
}

The following listing shows the assembly generated by the compiler:

\lstinputlisting[
        basicstyle=\tiny,
        numberstyle=\tiny,
        linewidth=1.0\textwidth,
        numbers=left,
        xleftmargin=0.03\textwidth
]{
code/vector_mult_real.asm
}

We briefly discuss the assembly. First, the zero register \texttt{xzr}
is copied into the loop counter register \texttt{x8} (line 1).
The \texttt{whilelo} instruction compares the counter status and the
length of the arrays, which is stored in register \texttt{x0}.
Relevant bits of the predication register \texttt{p1} are set to \emph{true}
 (line 2). The \texttt{ptrue} instruction sets all bits
in the predication register \texttt{p0} to \emph{true} (line 3).

In the loop body, the predicated \texttt{ld1d} instructions load slices of
the arrays $x$ and $y$ into the vector registers \texttt{z0} and
\texttt{z1}, respectively (lines 5--6). Inactive vector elements
are set to zero in the target registers, as indicated by \texttt{p1/z}.
All vector elements are multiplied pairwise using the
unpredicated instruction \texttt{fmul} (line 7).
The predicated store instruction \texttt{st1d} writes the
active elements of the result vector to memory (line 8).
The loop counter register \texttt{x8} is incremented by the SVE vector
length (in \texttt{double}) (line 9). The predication for the next
loop iteration is assembled (lines 10--12). The loop is iterated until all
array elements are processed.

It is important to note that the SVE vector length
does not appear explicitly. The number of loop iterations is determined
by the vector length implemented in the hardware. Predicated operations
eliminate the need for tail recursion, which is required on some other SIMD
architectures if the last remaining fraction of the data do not fit
exactly into the vector registers.

% -------------------------------------------------------------------------------------------------
\subsection{Complex Arithmetics}
\label{sec:cppcodecplxarith}
% -------------------------------------------------------------------------------------------------

As an example of complex arithmetics we consider pairwise
multiplication of complex array elements, with the result being stored in
a third array. The following code shows the C++ implementation of the
operation $z_i \leftarrow x_i \times y_i$ for arrays of the data type
\texttt{std::complex<double>} with $n$ elements each:

\lstinputlisting[
        language=C++,
        basicstyle=\tiny,
        numberstyle=\tiny,
        linewidth=1.0\textwidth,
        numbers=left,
        xleftmargin=0.03\textwidth
]{
code/vector_mult_cplx.cc
}

The following listing shows the assembly generated by the compiler:

\lstinputlisting[
        basicstyle=\tiny,
        numberstyle=\tiny,
        linewidth=1.0\textwidth,
        numbers=left,
        xleftmargin=0.03\textwidth
]{
code/vector_mult_cplx.asm
}

We briefly discuss the assembly, focusing on the differences to
multiplication of real array elements shown in
Section~\ref{sec:cppcoderealarith}.

First, the loop counter and the predication registers are initialized
(lines 1--3). In the loop body, the predicated structure load instructions
\texttt{ld2d} load slices of the two-element arrays $x$ and $y$ into four
vector registers. The real parts of the arrays $x$ and $y$ are loaded into
the vector registers \texttt{z0} and \texttt{z2}, respectively (line 6).
The imaginary parts of the arrays $x$ and $y$ are loaded into the
vector registers \texttt{z1} and \texttt{z3}, respectively (line 7).
Processing continues with real arithmetics, including multiplication,
multiply-add, and multiply-subtract (lines 10--15). The real parts of the
result are stored in vector register \texttt{z6}. The
imaginary parts of the result are stored in vector register
\texttt{z7}. The result vectors are written back to memory using the
predicated structure store instruction \texttt{st2d}. This instruction
reassembles two-element structures from two vector registers and
writes them into contiguous memory (line 16). The loop body is iterated
until all array elements are processed.

We note that the ARM SVE compiler generates assembly using structure
load/store. Real arithmetics is used for data processing.
The compiler does not exploit the full SVE ISA, which comprises
dedicated instructions for complex arithmetics. The reason is the
lack of support for complex arithmetics in the LLVM 5 backend of the
compiler.

% -------------------------------------------------------------------------------------------------
\subsection{Complex Arithmetics using SVE ACLE (I)}
\label{sec:cppcodecplxarithacle}
% -------------------------------------------------------------------------------------------------

As an example of complex arithmetics using SVE ACLE we consider pairwise
complex multiplication of arrays of complex numbers, with the result being
stored in a third array. In this example we implement complex numbers in
arrays of \texttt{double} with $2n$ elements each. Real and imaginary
parts of the array are interleaved $(re_1, im_1, re_2, im_2, \ldots)$.
We note that this implementation is equivalent to using arrays with
$n$ elements of \texttt{std::complex<double>} each. The following code
shows the C++ implementation of the operation
$z_i \leftarrow x_i \times y_i$ using SVE ACLE:

\lstinputlisting[
        language=C++,
        basicstyle=\tiny,
        numberstyle=\tiny,
        linewidth=1.0\textwidth,
        numbers=left,
        xleftmargin=0.03\textwidth
]{
code/vector_mult_cplx_acle.cc
}

We briefly discuss the details of this implementation. At first the
predication \texttt{pg} and SVE ACLE data types are declared (lines 2--4).
We use the \texttt{for}-loop for our implementation of complex
multiplication. The loop counter is incremented
after each loop iteration calling \texttt{svcntd()} (line 6). This
intrinsic function returns the SVE vector register length (in \texttt{double}).
In the loop body, we use the intrinsic function \texttt{svld1()} to load
slices of the arrays $x$ and $y$ without decomposing the array elements
(lines 8--9). Computation proceeds with multiply-add of complex
numbers using two calls to \texttt{svcmla()} (the intrinsic function
for the \texttt{FCMLA} instruction introduced in Section~\ref{sec:complexarith})
(lines 10--11). The first vector operand of the first \texttt{FCMLA}
instruction consists of zeros, resulting in complex multiplication
adding zero. The result vector is stored back into contiguous
memory using the predicated \texttt{svst1()} function (line 12). The
loop body is iterated until all data are processed.

The following listing shows the assembly generated by the compiler:

\lstinputlisting[
        basicstyle=\tiny,
        numberstyle=\tiny,
        linewidth=1.0\textwidth,
        numbers=left,
        xleftmargin=0.03\textwidth
]{
code/vector_mult_cplx_acle.asm
}

We briefly discuss the assembly. All function calls to SVE ACLE
instrinsic functions in the C++ code are directly translated into
assembly. No additional SVE instructions are generated. We conclude
that hardware support for complex arithmetic is accessible by
using SVE ACLE.

% -------------------------------------------------------------------------------------------------
\subsection{Complex Arithmetics using SVE ACLE (II)}
\label{sec:cppcodecplxarithaclenoloop}
% -------------------------------------------------------------------------------------------------

As the last example of complex arithmetics using SVE ACLE we again
consider pairwise complex multiplication of arrays of complex numbers,
with the result being stored in a third array. This example is
almost identical to Section~\ref{sec:cppcodecplxarithacle}.
However, here we omit the \texttt{for}-loop and use
the full SVE vector length implemented in the hardware for computation.
This implementation mimics programming for fixed-size SIMD registers
and is eminently suitable for small arrays of the size of vector registers.
The following code shows the C++ implementation of the operation
$z_i \leftarrow x_i \times y_i$ using SVE ACLE:

\lstinputlisting[
        language=C++,
        basicstyle=\tiny,
        numberstyle=\tiny,
        linewidth=1.0\textwidth,
        numbers=left,
        xleftmargin=0.03\textwidth
]{
code/vector_mult_cplx_acle_no_loop.cc
}

The following listing shows the assembly generated by the compiler:

\lstinputlisting[
        basicstyle=\tiny,
        numberstyle=\tiny,
        linewidth=1.0\textwidth,
        numbers=left,
        xleftmargin=0.03\textwidth
]{
code/vector_mult_cplx_acle_no_loop.asm
}

We conclude that for small arrays of the size of the SVE vector length
it is possible to omit the loop overhead implied by the VLA
programming model. We note that the resulting binaries will only be
operating correctly on matching SVE hardware.

%%% Local Variables:
%%% mode: latex
%%% TeX-master: "qpace4"
%%% End:

%% file: section_sve_enabling_grid.tex
% ==================================================================================================
\section{SVE-enabling Grid}\label{sec:sve-enabling-grid}
% ==================================================================================================

% --------------------------------------------------------------------------------------------------
\subsection{Strategies for enabling SVE in Grid}
\label{sec:proposals}
% --------------------------------------------------------------------------------------------------

As described in Section \ref{sec:dlayout}, Grid adapts the data layout
to the available vector length. Hence we have to set a vector length at
compile time, despite SVE being vector-length agnostic.

To enable SVE optimizations within Grid we have two options.
First, we can use Grid's generic implementation without any
architecture-specific optimizations, relying on the auto-vectorization
capabilities of the armclang compiler. Second, we add a SVE-specific
implementation to Grid's lower-level abstraction layer described in
Section \ref{sec:grid}. Current compiler heuristics are not good
enough to generate SVE instructions for complex arithmetic, as shown in
Section \ref{sec:cppcodecplxarith}. Therefore we decided to use ACLE
to enable hardware support for complex arithmetics.

The core of Grid's abstraction layer is a template class that enables
direct access to vector registers using intrinsic
data types. These data types are declared as member data. An example is
\texttt{\_\_m512d}, which defines a vector of 8 double-precision floating-point
numbers on AVX-512 architectures. This implementation scheme is feasible due to the
compiler's capability of auto-generating loads (stores) from (to) the
intrinsic data types.

For SVE this kind of implementation is not feasible because sizeless data types
cannot be used as class member data. Therefore we use ordinary arrays as
class member data and implement SVE ACLE only for data processing within functions.
Data processing was optimized for arrays of the size of the vector registers.
An example of how this is implemented was shown in Section \ref{sec:cppcodecplxarithaclenoloop}.

% --------------------------------------------------------------------------------------------------
\subsection{Implementation Details}
% --------------------------------------------------------------------------------------------------

As proposed in the last section we do not follow the VLA programming
model. Instead, our implementation is bound to the vector length
of the target hardware. Therefore, the Grid binaries are not necessarily
portable across different platforms. However, our SVE implementation
of Grid is portable at the cost of full compilation of the code base.
This is not a problem since the compilation time of Grid is insignificant compared to the time needed to perform
LQCD simulations.

To fix the SVE implementation of Grid to the target hardware we introduce
the compile-time constant \texttt{SVE\_VECTOR\_LENGTH},
which represents the SVE vector length in bytes. At the time of
writing this contribution SVE is enabled in Grid for 128-bit,
256-bit, and 512-bit vector implementations. We
note that at present Grid only supports up to 512-bit
architectures. Wider vectors (e.g., 1024 bit), are possible but
specialization of some of the lower-level functionality is necessary.

We introduce a templated C++ structure
\texttt{vec<T>}
with (aligned) ordinary array \texttt{v} as member data.
By definition the array is always of the size of the SVE vector length,
irrespective of the data type in use.
Specializations of the template typename \texttt{T} support
64-bit, 32-bit, 16-bit floating-point numbers and 32-bit integers.
Grid does not support calculations using 16-bit floating-point
numbers. This data type is used only for data compression upon data
exchange over the communications network.

We exploit different features of \cite{acle}, which we augmented
by the utility C++ templated structure \texttt{acle<T>}.
It is used to simplify mapping C++ data types in Grid to data types
supported by SVE ACLE. It is also used to provide various
definitions for predication.

% --------------------------------------------------------------------------------------------------
\subsection{Code Example}
% --------------------------------------------------------------------------------------------------

% SVE example
Complex multiplication is implemented as a templated C++ function of the
\texttt{MultComplex} structure. We use the vector-length specific
implementation introduced in Section \ref{sec:cppcodecplxarithaclenoloop}.
The following listing shows how complex multiplication is implemented
in the SVE-enabled version of Grid:

% Nils' cleaned up SVE implementation, as of July 2018
\lstinputlisting[
        language=C++,
        basicstyle=\tiny,
        numberstyle=\tiny,
        linewidth=1.0\textwidth,
        numbers=left,
        xleftmargin=0.03\textwidth
]{
code/grid_sve_cmul.cc
}

% --------------------------------------------------------------------------------------------------
\subsection{Implementation Verification}\label{sec:verification}
% --------------------------------------------------------------------------------------------------

% Testing
Grid implements about 100 ready-made tests and benchmarks. We have selected
40 representative tests and benchmarks for verification of the SVE-enabled
version of Grid for different SVE vector lengths using the ARM clang
18.3 compiler and the ARM SVE instruction emulator ArmIE 18.1. The SVE
emulator allows for functional verification of the SVE code generated by
the compiler. It also allows for defining the SVE vector length as
a command line parameter.

The majority of tests and benchmarks complete with success. However, some
tests fail due to incorrect results for some choices of the SVE vector length
and implementations of the predication. We attribute the failing tests to
minor issues of the ARM SVE toolchain, which is still under development.

% --------------------------------------------------------------------------------------------------
\subsection{Alternative Implementation of Complex Arithmetics}
% --------------------------------------------------------------------------------------------------

The silicon provider determines the SVE vector length and also
the performance characteristics of the hardware. The performance
signatures of the instructions might differ across different SVE
platforms. It is not guaranteed that the \texttt{FCMLA} instruction
outperforms alternative implementations of complex arithmetics.
Therefore, we have also implemented complex arithmetics based on
instructions for real arithmetics at the cost of higher instruction
count and cutting down on the effectiveness of SVE vector register
usage.

%%% Local Variables:
%%% mode: latex
%%% TeX-master: "qpace4"
%%% End:

%% file: section_related_work.tex
% ==================================================================================================
\section{Related Work}\label{sec:related-work}
% ==================================================================================================

Significant efforts have been made in the past by the LQCD community to provide
domain-specific libraries optimized for specific architectures.
One example is the QPhiX library \cite{QPhiX} that was specifically designed for Intel's Xeon Phi
architecture.
A more general approach targeting various x86 SIMD ISAs, but in particular AVX-512,
is Grid \cite{2015arXiv151203487B}.
Meanwhile exploratory studies have been performed to extend the portability of Grid to other types of
architectures, including GPU-accelerated ones \cite{Boyle:2017gzg}.
A  much earlier effort targeting architectures comprising NVIDIA GPUs supporting CUDA
resulted in the QUDA library \cite{Clark:2009wm},
which has meanwhile been used for several generations of GPU-accelerated supercomputers.

Little work has  been published so far on the evaluation of SVE for scientific computing
applications, including LQCD.
Some earlier work in this context has been published at last year's edition of CLUSTER.
In \cite{8049003} the authors report on performance results for selected numerical kernels
that were generated using a gem5 simulation setup.

%%% Local Variables:
%%% mode: latex
%%% TeX-master: "qpace4"
%%% End:

%% file: section_conclusion.tex
% ==================================================================================================
\section{Conclusion and Future Work}\label{sec:conclusion}
% ==================================================================================================

In this contribution we provided a brief introduction into the pertinent features of
applications for LQCD simulations. Focussing on the main computational task we
showed how these applications could benefit from the ARM ISA extension SVE.
By enabling the LQCD community code Grid for SVE,

% equalize columns
\newpage

\noindent we could explore how well this
new ISA can be exploited.

The results are very promising. The source code
is available \cite{sources}.

At the time of writing this contribution, it is not yet possible
to perform a reliable
assessment of the performance of the SVE-enabled version of Grid due to the lack of
processor architectures supporting the SVE architecture or simulators for such architectures.

%%% Local Variables:
%%% mode: latex
%%% TeX-master: "qpace4"
%%% End:

%% file: qpace4.bbl
\begin{thebibliography}{10}
\providecommand{\url}[1]{#1}
\csname url@samestyle\endcsname
\providecommand{\newblock}{\relax}
\providecommand{\bibinfo}[2]{#2}
\providecommand{\BIBentrySTDinterwordspacing}{\spaceskip=0pt\relax}
\providecommand{\BIBentryALTinterwordstretchfactor}{4}
\providecommand{\BIBentryALTinterwordspacing}{\spaceskip=\fontdimen2\font plus
\BIBentryALTinterwordstretchfactor\fontdimen3\font minus
  \fontdimen4\font\relax}
\providecommand{\BIBforeignlanguage}[2]{{%
\expandafter\ifx\csname l@#1\endcsname\relax
\typeout{** WARNING: IEEEtran.bst: No hyphenation pattern has been}%
\typeout{** loaded for the language `#1'. Using the pattern for}%
\typeout{** the default language instead.}%
\else
\language=\csname l@#1\endcsname
\fi
#2}}
\providecommand{\BIBdecl}{\relax}
\BIBdecl

\bibitem{PhysRevD.10.2445}
K.~G. Wilson, ``Confinement of quarks,'' \emph{Phys. Rev. D}, vol.~10, pp.
  2445--2459, 1974.

\bibitem{Edwards:2004sx}
R.~G. Edwards and B.~Joo, ``{The Chroma software system for lattice QCD},''
  \emph{Nucl. Phys. Proc. Suppl.}, vol. 140, p. 832, 2005.

\bibitem{Clark:2009wm}
M.~A. Clark, R.~Babich, K.~Barros, R.~C. Brower, and C.~Rebbi, ``{Solving
  Lattice QCD systems of equations using mixed precision solvers on GPUs},''
  \emph{Comput. Phys. Commun.}, vol. 181, pp. 1517--1528, 2010.

\bibitem{2015arXiv151203487B}
P.~{Boyle}, A.~{Yamaguchi}, G.~{Cossu}, and A.~{Portelli}, ``{Grid: A next
  generation data parallel C++ QCD library},'' \emph{ArXiv:1512.03487}, 2015.

\bibitem{7924233}
N.~Stephens, S.~Biles, M.~Boettcher, J.~Eapen, M.~Eyole, G.~Gabrielli,
  M.~Horsnell, G.~Magklis, A.~Martinez, N.~Premillieu, A.~Reid, A.~Rico, and
  P.~Walker, ``{The ARM Scalable Vector Extension},'' \emph{IEEE Micro},
  vol.~37, no.~2, pp. 26--39, 2017.

\bibitem{acle}
\BIBentryALTinterwordspacing
{ARM}, ``{ARM C Language Extensions for SVE},'' Tech. Rep., 2017. [Online].
  Available:
  \url{https://developer.arm.com/docs/100987/latest/arm-c-language-extensions-for-sve}
\BIBentrySTDinterwordspacing

\bibitem{QPhiX}
\BIBentryALTinterwordspacing
{QPhiX library}. [Online]. Available:
  \url{https://github.com/JeffersonLab/qphix}
\BIBentrySTDinterwordspacing

\bibitem{Boyle:2017gzg}
P.~A. Boyle, M.~A. Clark, C.~DeTar, M.~Lin, V.~Rana, and A.~V. Avil\'es-Casco,
  ``{Performance Portability Strategies for Grid C++ Expression Templates},''
  \emph{EPJ Web Conf.}, vol. 175, p. 09006, 2018.

\bibitem{8049003}
{Y. Kodama, T. Odajima, M. Matsuda, M. Tsuji, J. Lee, and M. Sato},
  ``{Preliminary Performance Evaluation of Application Kernels Using ARM SVE
  with Multiple Vector Lengths},'' in \emph{{2017 IEEE International Conference
  on Cluster Computing (CLUSTER)}}, {2017}, pp. {677--684}.

\bibitem{sources}
\BIBentryALTinterwordspacing
{SVE-enabled Grid}. [Online]. Available:
  \url{https://github.com/nmeyer-ur/Grid/tree/feature/arm-sve}
\BIBentrySTDinterwordspacing

\end{thebibliography}
